\documentclass[lettersize,journal]{IEEEtran}

\usepackage{subfigure}
\usepackage{graphicx}
\usepackage{amsmath}

\usepackage{amssymb}
\usepackage{bm}
\usepackage{float}
\usepackage{booktabs}
\usepackage{multirow}
\usepackage{makecell} 
\usepackage{verbatim}

\usepackage{color}
\usepackage{soul}
\soulregister\cite7
\soulregister\ref7
\makeatletter
\newcount\SOUL@minus %% <- without this line the page number would be reset to 0
\makeatother

\begin{document}
\bstctlcite{MyBSTcontrol}

\title{Domain Adversarial Graph Convolutional Network Based on RSSI and Crowdsensing for Indoor Localization}

\author{Mingxin Zhang, Zipei Fan, Ryosuke Shibasaki, Xuan Song
\thanks{This work was supported by the Ministry of Education, Culture, Sports, Science, and Technology under Grant 22H03573. \textit{(Corresponding author: Zipei Fan.)}}
\thanks{Mingxin Zhang, Zipei Fan, Ryosuke Shibasaki, and Xuan Song are with the Center for Spatial Information Science, The University of Tokyo, Chiba 277-8568, Japan (e-mail: zhangmx@csis.u-tokyo.ac.jp, fanzipei@iis.u-tokyo.ac.jp, shiba@csis.u-tokyo.ac.jp, songxuan@csis.u-tokyo.ac.jp).}}

%\markboth{Journal of \LaTeX\ Class Files,~Vol.~XX, No.~X, August~2021}%


\maketitle

\begin{abstract}
In recent years, the use of WiFi fingerprints for indoor positioning has grown in popularity, largely due to the widespread availability of WiFi and the proliferation of mobile communication devices. However, many existing methods for constructing fingerprint datasets rely on labor-intensive and time-consuming processes of collecting large amounts of data. Additionally, these methods often focus on ideal laboratory environments, rather than considering the practical challenges of large multi-floor buildings. To address these issues, we present a novel WiDAGCN model that can be trained using a small number of labeled site survey data and large amounts of unlabeled crowdsensed WiFi fingerprints. By constructing heterogeneous graphs based on received signal strength indicators (RSSIs) between waypoints and WiFi access points (APs), our model is able to effectively capture the topological structure of the data. We also incorporate graph convolutional networks (GCNs) to extract graph-level embeddings, a feature that has been largely overlooked in previous WiFi indoor localization studies. To deal with the challenges of large amounts of unlabeled data and multiple data domains, we employ a semi-supervised domain adversarial training scheme to effectively utilize unlabeled data and align the data distributions across domains. Our system is evaluated using a public indoor localization dataset that includes multiple buildings, and the results show that it performs competitively in terms of localization accuracy in large buildings.
\end{abstract}


\section{Introduction}
Accurate indoor localization technology has the potential to support a variety of smart applications that involve sensing the environment and improving user experiences. While GPS is the dominant technology for outdoor localization, it tends to be less accurate in indoor environments due to the lack of outdoor exposure. As a result, researchers have been exploring various techniques to develop reliable indoor location systems. In recent years, using existing WiFi infrastructure to build a localization system has emerged as a cost-effective and convenient solution, given the widespread availability of indoor WiFi signals \cite{li2021dafi,wu2017gain,10.1007/978-3-030-10997-4_32}. One approach that has gained particular attention is the use of Received Signal Strength Indication (RSSI) to build WiFi fingerprints, as it is easily obtainable from ubiquitous smartphone devices and does not require the use of indicators like Channel State Information (CSI) \cite{hernandez2020performing}.

Collecting a reliable dataset of WiFi signals through a site survey can be a labor-intensive and time-consuming process, particularly in large multi-level buildings. Although there have been some studies working on using labeled or unlabeled large dataset collected from users' devices \cite{wang2012no}, this also poses the risk of label contamination and the difficulty of utilizing unlabeled data. In addition, RSSIs can be affected by reflection and diffraction constantly happening in the indoor environment, becoming susceptible, particularly in large buildings. A more complex problem is, that in a multilevel building, signals from an AP can be received by devices on multiple floors, this may cause a single 2D coordinate label corresponding to multiple different RSSI features. 
\IEEEpubidadjcol
In this paper, we propose two models for indoor localization using WiFi signals: a semi-supervised WiFi Domain Adversarial Graph Convolutional Network (WiDAGCN) model and a supervised WiFi Attention Graph Convolutional Network (WiAGCN) model, both based on graph convolutional network (GCN) \cite{DBLP:conf/iclr/KipfW17}. To reduce the workload of data collection, our system utilizes crowdsensed data collected from user devices without the need for manual labeling. Because it is difficult to train a model with a small labeled dataset, it is necessary to apply the unlabeled crowdsensing data to the training process as well. Based on the semi-supervised adversarial training process, we conduct domain adaptation to align the data distribution of different data domains, making full use of unlabeled data and enhancing the robustness of the model. 

To preserve the information contained in the data, we create topology graphs based on the connections between access points (APs) and waypoints as the input feature, rather than using fingerprint vectors directly. This approach not only reduces the reliance on labels, but also allows the model to learn from the higher-order neighbors and improve its learning ability. Since seldom studies focus on the importance of the permutation invariance \cite{zaheer_deep_2017} in indoor localization problems, we also consider the importance of permutation invariance in indoor localization problems, and demonstrate the effectiveness of permutation invariant models in improving accuracy. The GCN model is known to have permutation invariance properties, making it a suitable choice for building a graph-based permutation invariant model \cite{Pei2020Geom-GCN:,pmlr-v100-liu20a}. Also, GCN can extract graph-level embeddings, which allows us to obtain complex relationships between data points that may not be easily captured by other types of networks. 

\begin{figure*}[h]
  \centering
  \includegraphics[width=\linewidth]{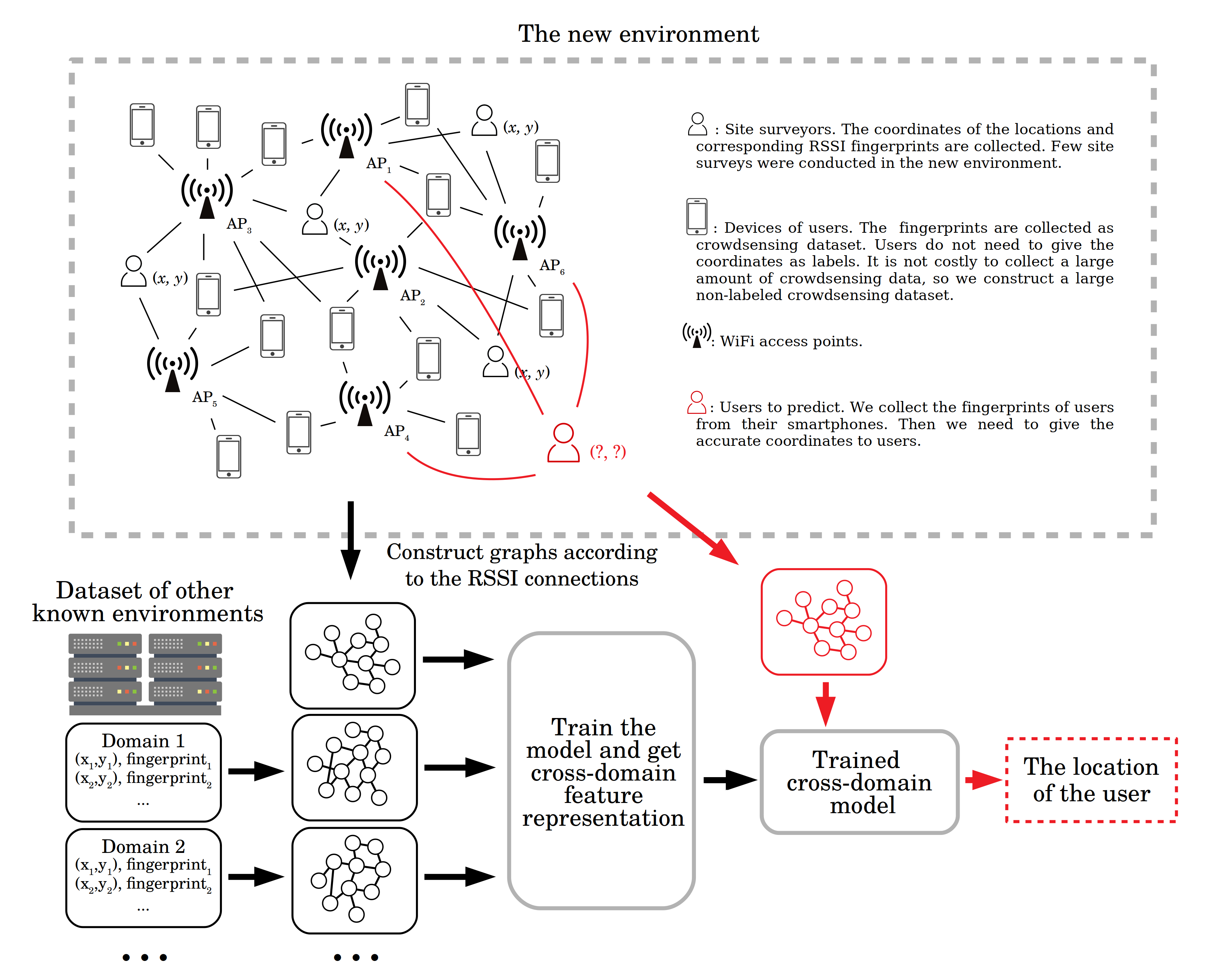}
  \caption{The Structure of the indoor localization system.}
  \label{fig:SystemStructure}
\end{figure*}

The structure of our indoor localization system is shown in Fig. \ref{fig:SystemStructure}. A small amount of labeled data is collected by site surveyors while a large unlabeled crowdsensing dataset is also built. This allows us to build a sufficiently rich and usable dataset with very little human effort. The model is trained using graphs from known environments, along with graphs from the new environment, allowing it to align the data distributions, improve robustness, and apply the knowledge learned from known environments to the new environment. The main contributions are summarized as follows.

\begin{itemize}
\item Our models are designed to be able to be trained with very few labeled data, meaning that large-scale site surveys are not necessary. Instead, unlabeled data collected from users can be utilized to aid the training process through the use of semi-supervised learning.

\item To make use of both labeled and unlabeled data, we present a new indoor localization system based on a semi-supervised graph convolutional network.

\item We test the performance of different models, and the result shows that permutation invariant features can indeed improve the accuracy of the indoor localization problem. We aggregate the whole subgraph features to get graph-level permutation invariant representations.

\item Previous graph-based approaches have not typically considered the application of indoor localization in complex real-world scenarios. Our model is able to achieve room-level positioning accuracy in large-scale multi-floor buildings using only a small amount of labeled data.  
\end{itemize}

The rest of the paper is organized as follows. We introduce some related work in Section \ref{sec:Related Work}, then we define the problem and describe the dataset preparation in Section \ref{sec:Preliminaries} and present the details of the model design and training method in Section \ref{sec:Methodology}. We set up the experiment and evaluate our models in Section \ref{sec:Experiment}. Finally, we discuss and conclude the work in Section \ref{sec:Discussion} and Section \ref{sec:Conclusion}.

\section{Related Work}
\label{sec:Related Work}

In this section, we provide a comprehensive review of various indoor localization technologies, with a specific focus on the localization system based on WiFi. We will also present some machine learning-based methods for WiFi localization and describe the application of GCN models in this context.

\subsection{Indoor Localization System}
In recent years, researchers have been actively investigating various technologies for indoor localization systems, with a focus on improving accuracy and robustness. In a previous study, the SnapLoc system, which leverages Ultra-wideband (UWB) technology to estimate positions using the time difference of arrival signals between anchors, was developed \cite{10.1145/3302506.3310389}. Other studies have focused on using embedded sensors on mobile phones, like the cellular, to provide localization services, such as the deep learning-based CellinDeep \cite{8570849}. Among the various options, WiFi-based localization methods have become increasingly popular due to their widespread availability. Different indicators, such as CSI \cite{hernandez2020performing} and mmWave WiFi \cite{feng2022survey}, have been used, with RSSIs being particularly easily obtainable from ubiquitous smartphone devices.

Since the pioneering study of RADAR \cite{bahl2000radar}, many studies have explored indoor localization methods using RSSI as location fingerprint \cite{wu2017gain, 10.1007/978-3-030-10997-4_32, 7248785, xu2018embracing, sugasaki2017robust}. With the development of machine learning technology, there are many novel ideas to solve the indoor localization problem.

\subsection{Machine Learning Based WiFi Localization}
W. Xue et al. proposed a weighted algorithm based on the physical distance of the RSSI \cite{8362651}, and compared its performance with various K-Nearest Neighbors (KNN) models. The model was tested on two floors of a building and achieved an accuracy of 5.03 m. J. Niu et al. proposed the WicLoc \cite{7248785} and achieved an average localization error of 4.60 meters in a single environment. They also used different size of the dataset to show the robustness of the model. However, their experimental environments were relatively ideal compared to real-world conditions. Traditional machine learning methods have faced challenges in improving performance further due to the high-dimensional data structure and complexity of indoor environments. Therefore, deep learning methods may be more suitable for addressing these types of problems \cite{feng2022survey}.

As early as 2002, researchers have been exploring the use of multi-layer fully connected neural networks for WiFi-based indoor localization using radio frequency (RF) data \cite{battiti2002location}. Since then, many other types of structures have been considered in an effort to extract richer features hidden in the data and improve performance in more complex experimental designs. Some stacked autoencoder (SAE) models based on fully connected networks have been tested, which can be trained in an unsupervised manner and then combined with a classifier for supervised training to create a localization model \cite{belmannoubi2019deep}. The use of convolutional neural networks (CNNs) as feature extractors has also been widely explored, with WiFi fingerprint vectors being input directly into 1D-CNNs \cite{hsieh2019deep} or transformed into 2D format and processed using 2D-CNNs, similar to the way image data is commonly processed \cite{jang2018indoor}. Some other researchers proposed models from the aspect of analysing the trajectory or the moving pattern of the users. They naturally used long-short term memory (LSTM) to get the temporal features and improved the performance with the help of the time series \cite{zhang2021indoor}. While different models have varied structures, most WiFi indoor localization systems in recent years have primarily used fingerprint vectors composed of RSSI data as model inputs. While fingerprint vectors can be easily built, they do have limitations.

\subsection{Graph Convolutional Network}
To preserve more complete information and features in the input data, it is worthwhile to consider other data structures that can better represent the relationship between waypoints and APs. In contrast to fingerprint vectors, graphs can represent the complex topological relationships between different entities, making the learning of features and information hidden within graphs a focus of research. GCN \cite{DBLP:conf/iclr/KipfW17} can effectively learn and utilize features hidden in graphs through the use of spatial or spectral convolution operations. W. Song et al. modeled dynamic online community user behaviors with a recurrent neural network (RNN) and context-dependent social influence with a graph-attention neural network (GAT) \cite{velickovic2018graph, 10.1145/3289600.3290989}. X. Han et al. proposed a graph convolution collaborative filtering model based on the attention model and obtained more excellent feature using attention model to highlight the relationship and interaction information \cite{10.1145/3404555.3404641, 10.1145/3264909}. While GCN can express the deep graph feature, it also guarantees the permutation invariance of the model \cite{Pei2020Geom-GCN:,pmlr-v100-liu20a}. There also have been studies working on the network model optimization, for example, W. Chiang et al. proposed Cluster-GCN \cite{10.1145/3292500.3330925}, made it easier to train a large-scale GCN. Researchers have constructed graphs for different ubiquitous computing scenarios, such as simulating the physical network topology \cite{10.1145/3411818}, retaining the spatio-temporal information \cite{10.1145/3264957, 10.1145/3495003}. 

There have been a number of studies that have implemented graph-based localization and mobility prediction systems \cite{chiou2020zero, 10.1145/3495003}, but to the best of our knowledge, only a few studies have focused on using graph neural networks to build WiFi indoor localization systems \cite{inproceedings, wu2022multi}.

In summary, current research on RSSI-based indoor positioning tends to focus on laboratory or simple and ideal environments, while solutions suitable for complex, large-scale application scenarios often require labor-intensive and time-consuming data collection efforts. Furthermore, GCNs are well-suited for preserving topological feature information and the permutation invariance of sets. However, there is a lack of research on the use of GCNs for indoor localization.

\section{Preliminaries}
\label{sec:Preliminaries}
In this section, we briefly describe the indoor localization problem and we introduce the public dataset used in this paper and the data preprocessing method.  
\subsection{Problem Definition}
The signal strengths of different APs are described by the RSSI fingerprint corresponding to each waypoint. For each waypoint, let {\itshape n} be the number of selected APs in the environment. Then we can obtain an RSSI vector as $x_{RSSI}=[RSSI_1, ..., RSSI_n]$, where $RSSI_i$ is the RSSI value received from the $i$-th AP. Similarly, let $x_{bssid}=[bssid_1, ..., bssid_n]$, where $bssid_i$ is the basic service set identifier (bssid) of the $i$-th AP. Because only a few site surveys were conducted in our experiment scenario, there are very few labeled data in the target domain, which means we can only get a large number of crowdsensing data $x_{RSSI}$ and $x_{bssid}$ without corresponding coordinates. In addition, the equipment and arrangement of APs in different buildings are completely different. As a result, the focus of this paper is on methods that make full use of both labeled and unlabeled data, helping the model fully learn the target domain data distribution. We also aim to extract domain-invariant features and transfer knowledge learned in known source domains to the unknown target domain using cross-domain features.

\subsection{Dataset Preparation}
\label{sec:Dataset Preparation}
In this paper, the Microsoft Indoor Location Competition 2.0 dataset \cite{shu2021indoor} is used. Five buildings are selected as pretraining sites, which contain known data to pretrain the model, and one building is designated as the target site, where the main focus of domain adaptation lies. The details of the domain setup and dataset split are described in Section \ref{sec:Experiment Setup}. To filter out low-quality signal connections, only the 50 connections with the highest RSSI strength are considered for each waypoint in this paper. The RSSI values are standardized by $RSSI = (RSSI_{origin} - \mu) / \sigma$, where $RSSI_{origin}$ represents the original values of RSSI, $\mu$ and $\sigma$ represent the mean and standard deviation of $RSSI_{origin}$ respectively, and $RSSI$ represents standardized values. The bssids of APs are also transformed into consecutive integer indices, from 0 to $n_{bssid}-1$.

\begin{figure*}[h]
\centering
\subfigure[]{
    \includegraphics[width=0.23\linewidth]{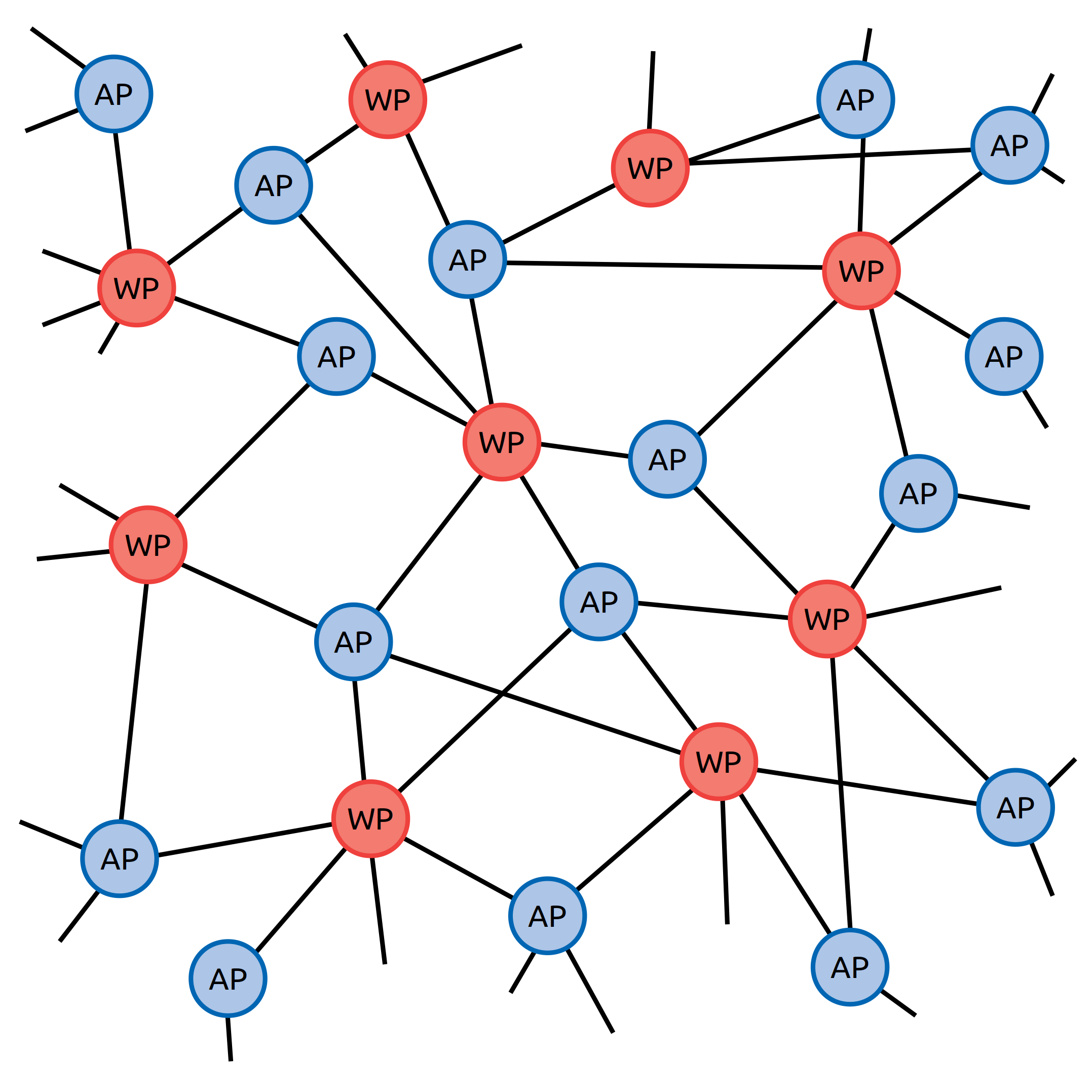}
}
\subfigure[]{
    \includegraphics[width=0.23\linewidth]{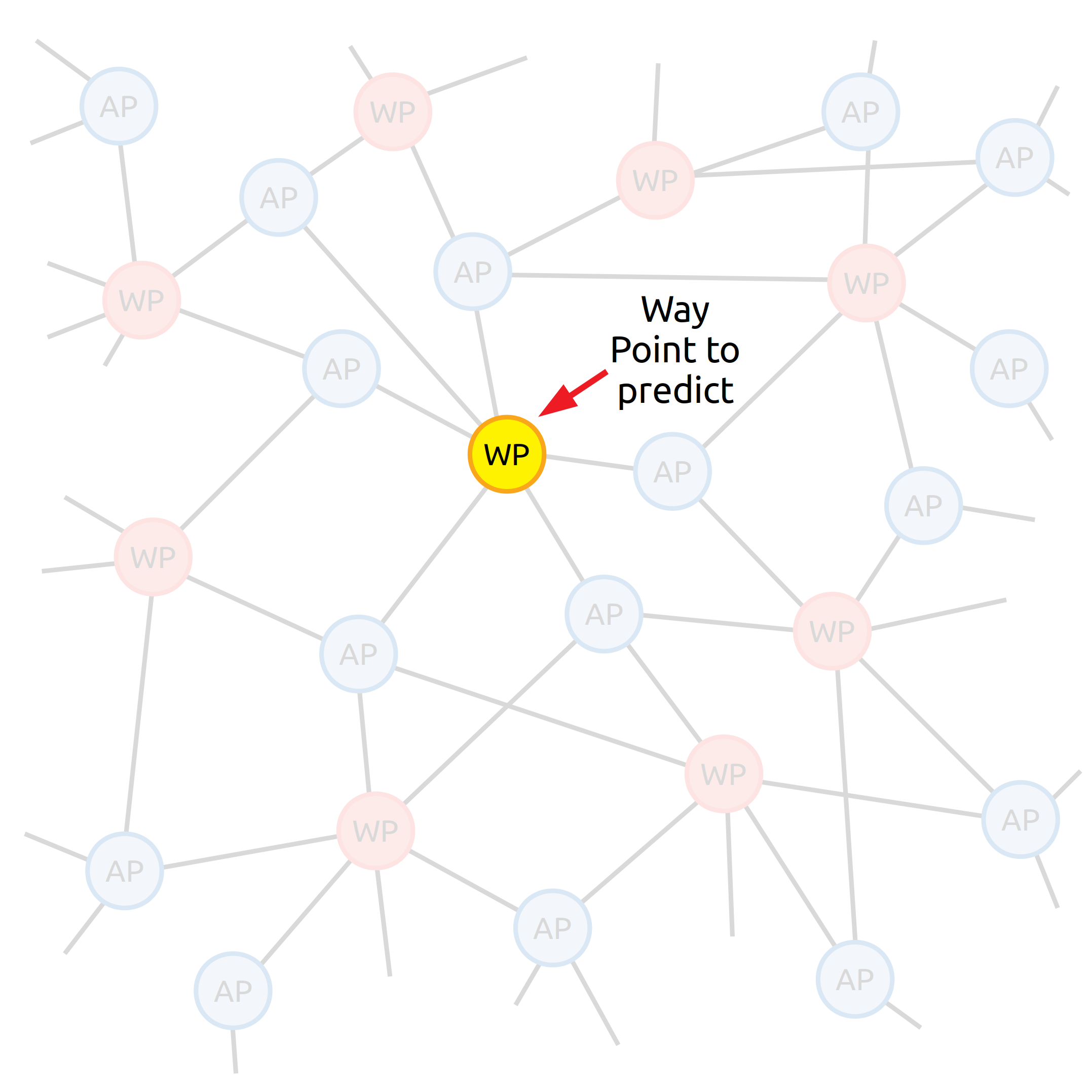}
}
\subfigure[]{
	\includegraphics[width=0.23\linewidth]{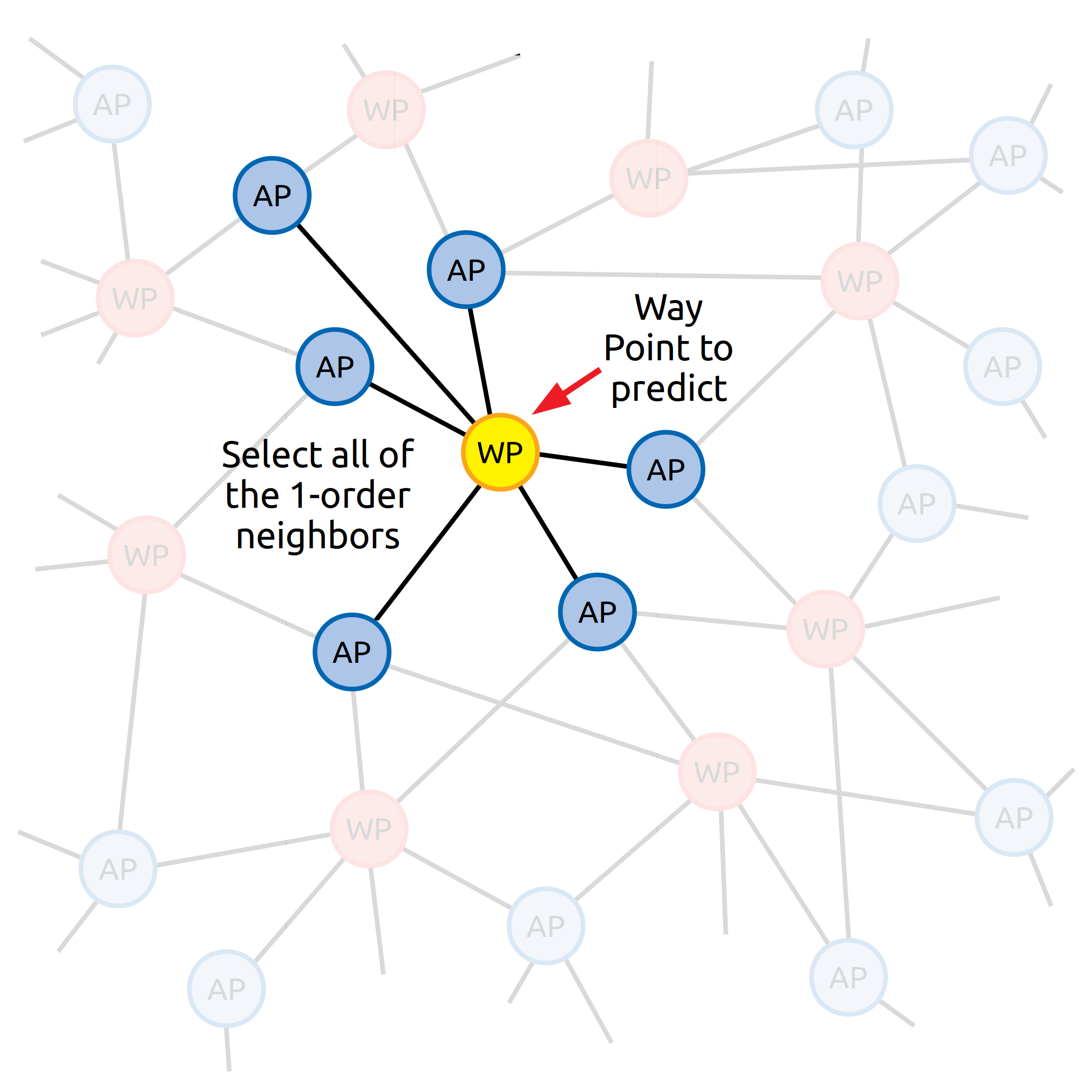}
}
\subfigure[]{
    \includegraphics[width=0.23\linewidth]{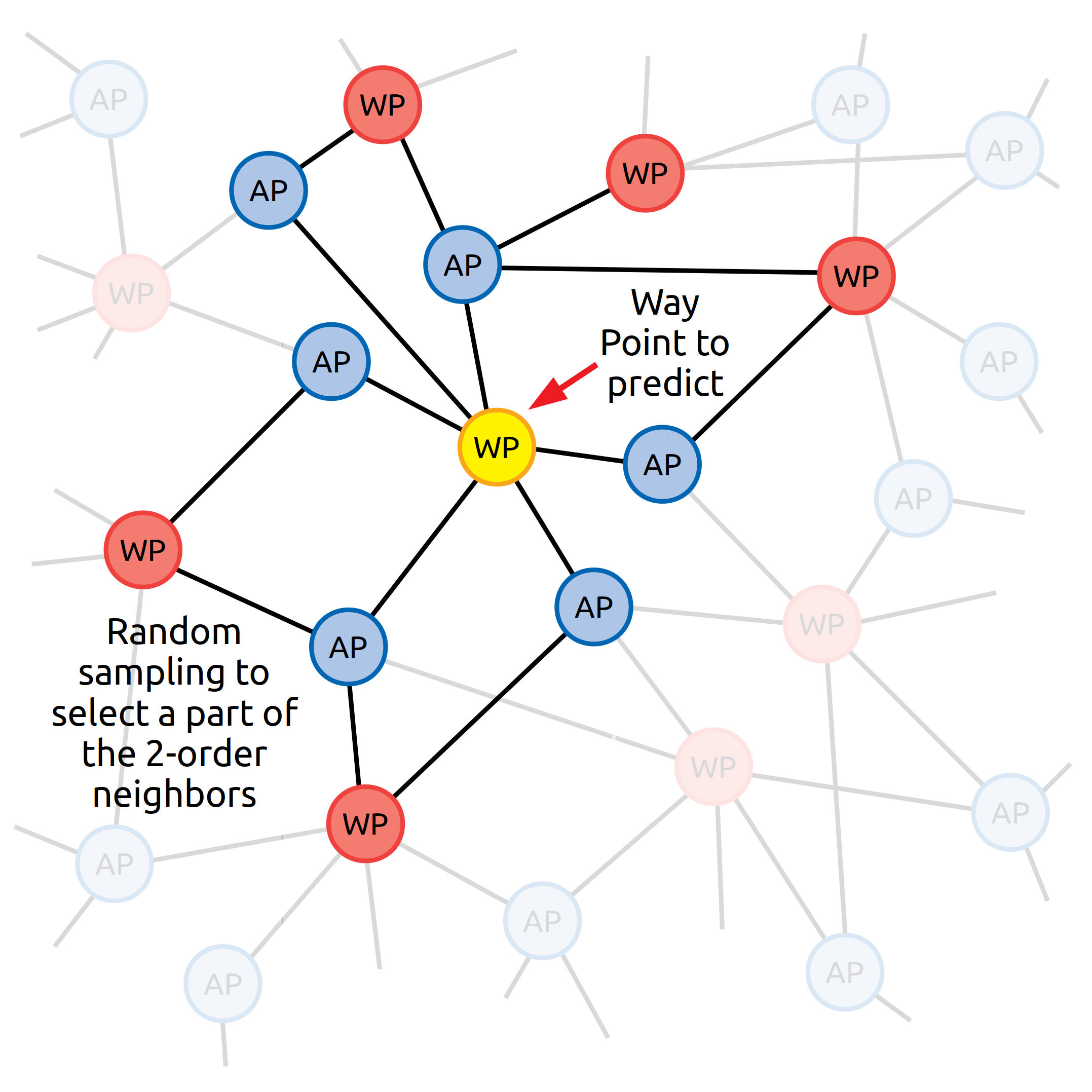}
}
\caption{The steps of subgraph construction. (a) Build the large graph. (b) Select the waypoint node to predict. (c) Select the 1st-order neighbors. (d) Select the 2nd-order neighbors.}
\label{fig:ConstructSubGraph}
\end{figure*}

\section{Methodology}
\label{sec:Methodology}
In this section, we will describe the graph structure and design of our proposed deep learning model, which serves as domain adaptation. And different training schemes will also be introduced.
\subsection{Graph Construction}
\label{sec:Graph Construction}
Through signal propagation between APs and devices, we can naturally obtain the connection relationship between APs and devices, thus establishing heterogeneous graphs. We represent the AP nodes as $V_{AP}$ and the waypoint nodes as $V_{WP}$. In this way, there are also two types of relationships between nodes, $E_{AP\rightarrow WP}$ and $E_{WP\rightarrow AP}$, representing the edges from APs to waypoints and from waypoints to APs respectively. Therefore, the node features can be represented as $F_{AP}=\{f^{AP}_1, f^{AP}_2, ...\}$ and $F_{WP}=\{f^{WP}_1, f^{WP}_2, ...\}$, the edge features can be represented as $F_{WP\rightarrow AP}=\{f^{WP\rightarrow AP}_1, f^{WP\rightarrow AP}_2, ...\}$ and $F_{AP\rightarrow\ WP}=\{f^{AP\rightarrow WP}_1, f^{AP\rightarrow WP}_2, ...\}$, where $f^{AP}_i$ represents the feature of the $i$-th AP, corresponding to the node $v^{AP}_i$. In this paper, we use the encoded bssids as the feature of AP nodes, which means that $f^{AP}_i=bssid_i$. And $f^{WP}_i$ is set as $(x_i,y_i)$, the 2D coordinate of the $i$-th waypoint, corresponding to $v^{WP}_i$. For the edges in the graph, they represent the relationship between APs and waypoints, so we can set $f^{WP\rightarrow AP}_i$ as the RSSI value between the starting point and the end point of the $i$-th edge $e^{WP\rightarrow AP}_i$. And the condition of $F_{AP\rightarrow WP}$ is the same. In this way, we define the heterogeneous graph as $G=(V,E,F_V,F_E)$, where
$
V=\{V_{WP}, V_{AP}\},\ 
E=\{E_{WP\rightarrow AP}, E_{AP\rightarrow WP}\},\ 
F_V=\{F_{WP}, F_{AP}\}
$
and
$
F_E=\{F_{WP\rightarrow AP}, F_{AP\rightarrow\ WP}\}
$
represent the nodes, edges and the features in the graph respectively.

To construct a large graph, we use all known nodes and edges. Inspired by the concept of Deep Sets, we use the aggregated features of the node to be predicted and its surrounding nodes rather than the features of a single node. The steps for constructing a subgraph are shown in Fig. \ref{fig:ConstructSubGraph}. A subgraph corresponding to each waypoint is created by traversing all waypoints. 

\begin{figure}[h]
  \centering
  \includegraphics[width=0.8\linewidth]{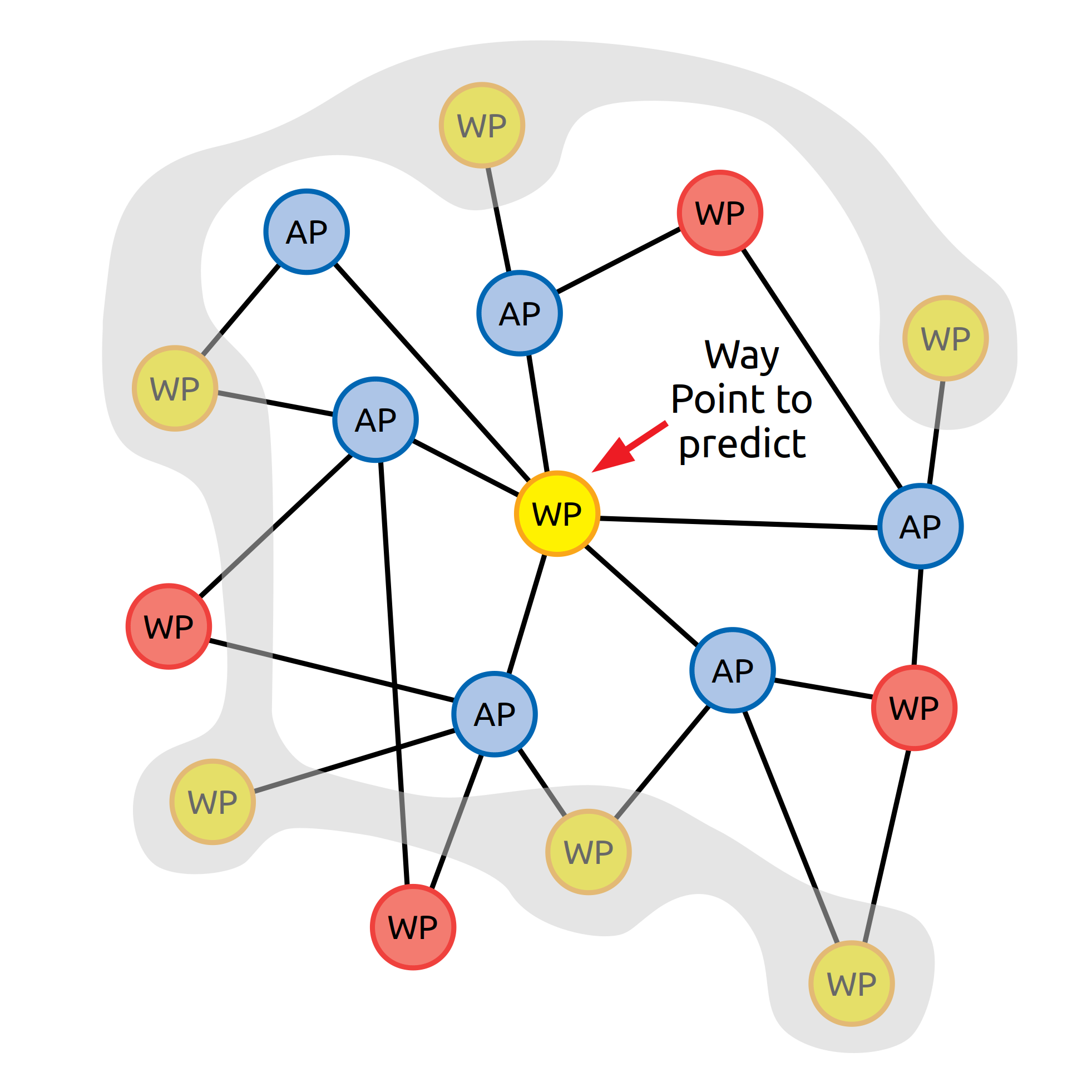}
  \caption{Subgraph with incomplete labels in target domain. The yellow waypoint nodes mean waypoints without coordinate features. The 2nd-order waypoint nodes without labels are marked with grey shade.}
  \label{fig:IncompletedSubGraph}
\end{figure}

In the case of the source domain, the feature is also the label of the waypoint, meaning that the features of the waypoints to be predicted are unknown. Therefore, we set an all-ones vector as the feature of the waypoint to be predicted (yellow node in Fig. \ref{fig:ConstructSubGraph}) and mark the waypoint to be predicted as a new type of node. We also create a pair of new edge types between the waypoint to be predicted and the AP (WP to predict $\rightarrow$ AP, AP $\rightarrow$ WP to predict). Next, all 50 neighbors of the waypoint to predict are selected and added to the subgraph. We randomly select five 2nd-order neighbors for each 1st-order neighbor and add them into the subgraph. There will be 2000 to 3000 nodes in a single subgraph if we use all 1st-order and 2nd-order neighbors, make the model difficult to train. Therefore, we use the method of multiple repeat sampling to reduce the complexity of the graph and to retain as much information as possible in the graph. We conduct 5 times repetitive sampling so there are 5 subgraphs with 301 nodes ($1 + 50 + 50 \times 5$) corresponding to each waypoint. All connections between the selected nodes are kept. The situation in the target domain is similar to that in the source domain, except that most of the waypoints in the target domain only have known fingerprint features and no labels. To handle this, we mark waypoints with unknown labels in the same way as waypoints to be predicted in the source domain. This results in subgraphs with incomplete labels, as shown in Fig. \ref{fig:IncompletedSubGraph}. The yellow waypoint nodes represent waypoints without coordinate features, and the 2nd-order waypoint nodes without labels are marked with a grey shade.

\begin{figure*}[h]
  \centering
  \includegraphics[width=\linewidth]{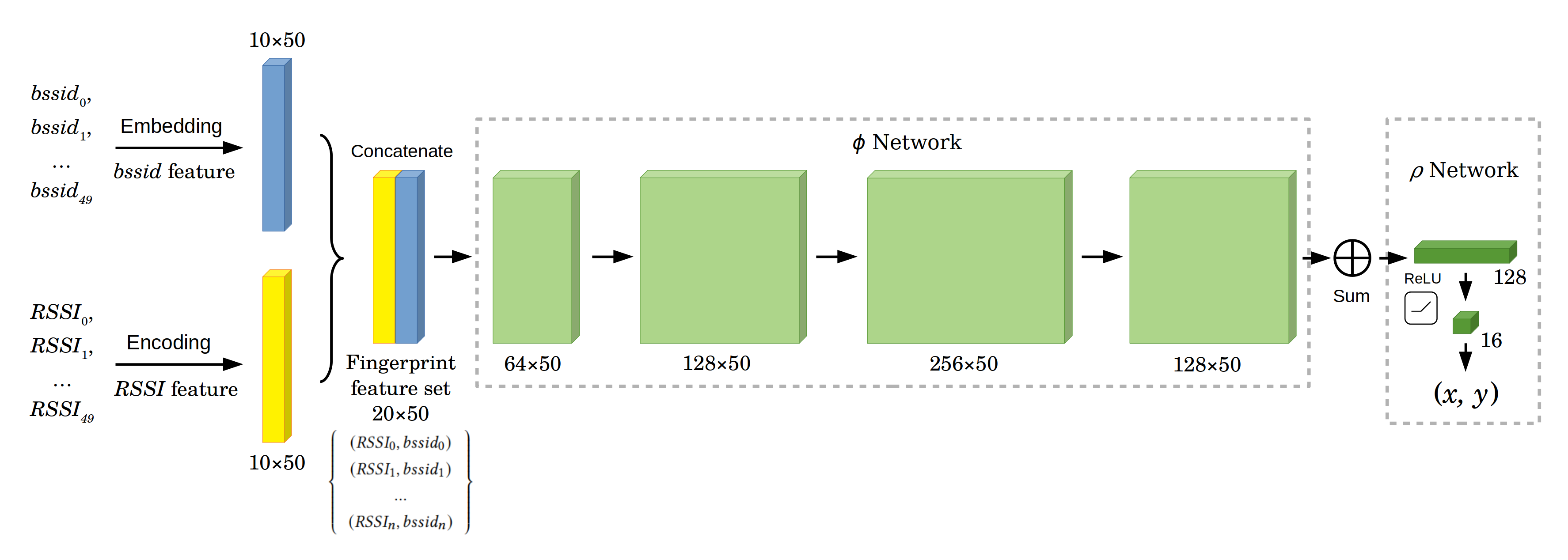}
  \caption{The structure of Deep Sets model. To verify whether permutation invariance has an important effect on the accuracy of indoor positioning, we designed this model inspired by Deep Sets and compared it with a common neural network without permutation invariant operation (the summation operation).}
  \label{fig:Deep Sets}
\end{figure*}

\subsection{Design of Neural Network}

\subsubsection{Deep Sets Model}
\label{sec:Deep Sets}
The model should only consider the values of the RSSI and the corresponding bssids, and not the order in which they appear. This property can be described as: The function $f$ acting on the set $X$ should be permutation invariant to the order of elements in $X$, i.e. for any permutation $\pi$, $f({x_1, ..., x_n})=f({x_{\pi(1)}, ..., x_{\pi(n)}})$. Previous research gave a theorem \cite{zaheer_deep_2017} of the function format: A function $f(X)$ operating on a set X having elements from a countable universe, is a valid set function. i.e., invariant to the permutation of instances in X, if and only if it can be decomposed in the form $\rho(\Sigma_{x\in X}\phi(x))$, for suitable transformation $\phi$ and $\rho$. Which permutation invariant operation to use is still a topic that needs to be discussed, here, in this theorem, the summation operation is chosen, so we also use the summation operation in this paper.

The idea of Deep Sets network is using the deep network to fit the function $\phi$ and $\rho$. The structure of the Deep Sets model implemented in this paper is shown in Fig. \ref {fig:Deep Sets}.

As shown in Fig. \ref {fig:Deep Sets}, elements of the fingerprint set are RSSI/bssid composite features corresponding to each available AP. For each waypoint, the model can be described as follows.

\begin{equation}
\begin{aligned}
X&=\left[
\begin{array}{cc} 
F_{RSSI}&F_{bssid}
\end{array}
\right]\\
&=
\left[
  \begin{array}{cc} 
    RSSI_0&bssid_0\\
    RSSI_1&bssid_1\\
    ...&...\\
    RSSI_{49}&bssid_{49}
  \end{array}
\right]
=\left[
  \begin{array}{c} 
    x_0\\
    x_1\\
    ...\\
    x_{49}
  \end{array}
\right],
\end{aligned}
\end{equation}

\begin{equation}
output=\rho(\Sigma_{x\in{X}}\phi(x))
=\rho(\phi(x_0)+\phi(x_1)+...+\phi(x_{49})).
\end{equation}

Because 50 WiFi signal connections are kept for each $(x, y)$ coordinate, there are 50 bssids and 50 corresponding RSSIs. The original bssids and RSSIs are processed by the embedding layer and the fully-connected encoding layer respectively, transformed into $10\times50$ feature matrices $F_{bssid}$ and $F_{RSSI}$. Then these two matrices are concatenated into a $20\times50$ matrix, representing the fingerprint feature set. After the $\phi$ network, the matrix is added along the axis that represents the set elements, resulting in a vector representing the features of the set. Finally, the $\rho$ network gives the predicted coordinate of the waypoint with the vector. To verify if the model retains the permutation invariant feature of the set and has a better performance, a fully-connected neural network with the same structure is constructed as a control group.

\subsubsection{WiDAGCN Model}
Inspired by the idea of the Deep Sets, we propose a WiFi Domain Adversarial Graph Convolutional Network (WiDAGCN). While GCN can express the deep graph feature, it also guarantees the permutation invariance of the model \cite{Pei2020Geom-GCN:,pmlr-v100-liu20a}. But first, before the description of WiDAGCN model, a supervised model that has a similar structure will be introduced.

\begin{figure*}[h]
  \centering
  \includegraphics[width=\linewidth]{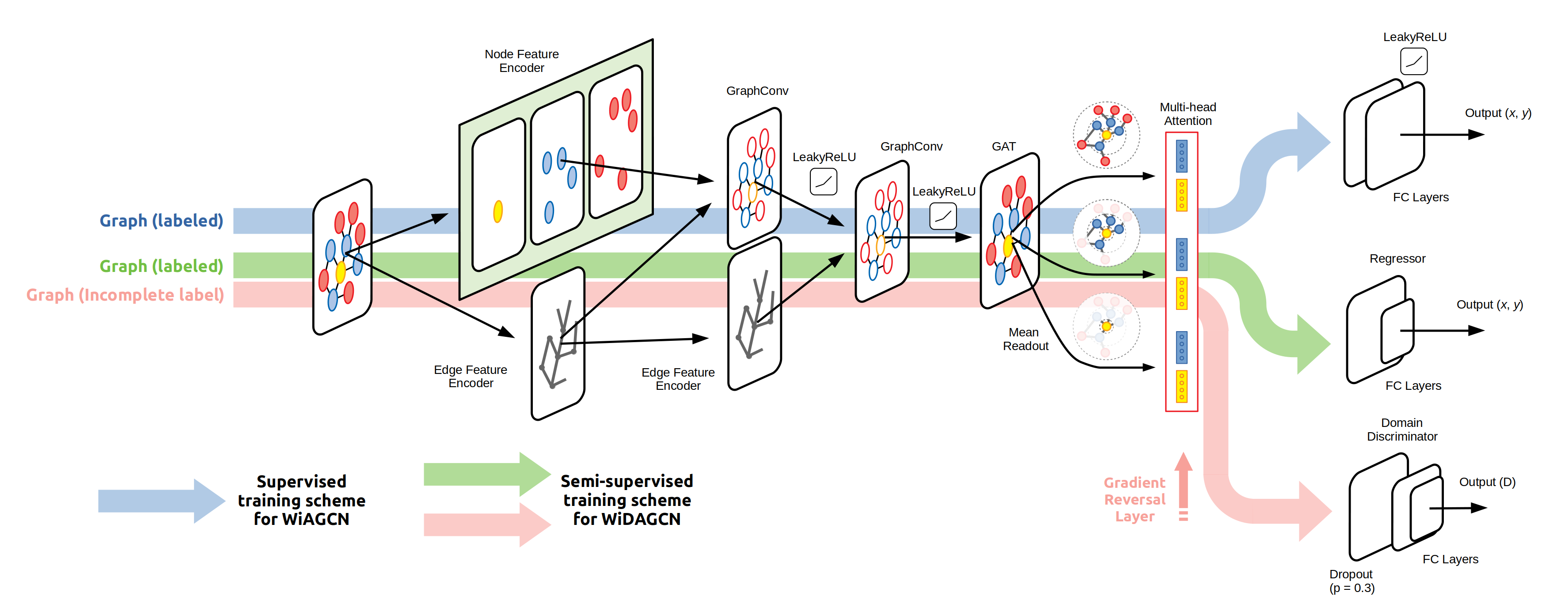}
  \caption{The structure of WiAGCN/WiDAGCN model. The part where the blue line passes through represents the structure of WiAGCN, and the part where the green line and pink line pass through represents the structure of WiDAGCN.}
  \label{fig:WiAGCN/WiDAGCN}
\end{figure*}

This supervised WiFi Attention Graph Convolutional Network (WiAGCN) model consists of signal encoding layers, GCN layers, and a graph-attention neural network (GAT) \cite{velickovic2018graph} layer. The structure of WiAGCN is shown in Fig. \ref{fig:WiAGCN/WiDAGCN}.

As mentioned in Section \ref{sec:Graph Construction}, the input subgraph contains three types of nodes and four types of edges. Similar to the Deep Sets model, an embedding layer is used to represent the bssid features, and other nodes and edge features are represented by full-connected layers. Two weighted heterogeneous graph convolutional layers are used to obtain the node embedding. With the GAT layer, we can get multiple node feature representations with the multi-head attention mechanism. The key point to maintaining the permutation invariant feature is the readout function, which is similar to the permutation invariant operation in Deep Sets model. We aggregate the mean node representation of three types of nodes and get a graph-level feature. Then we aggregate the central waypoint (the waypoint to predict) and the surrounding 1st-order neighbors, the APs, to get a local feature. Finally, we extract the feature of the center point. These three different level features are concatenated to obtain the final regression result. 

In this study, we use Deep Graph Library (DGL) \cite{https://doi.org/10.48550/arxiv.1909.01315} to implement the graph-based model. The principles of the graph neural network model used in DGL will be described next.

Because we use connections between APs and waypoints as edges and RSSI values as edge features, we should consider the weighted graph convolution to utilize the edge information. The weighted graph convolution used in DGL GraphConv module was introduced in the study \cite{DBLP:conf/iclr/KipfW17} and can be mathematically defined as:
\begin{equation}
h^{(l+1)}_i=\sigma(b^{(l)}+\sum_{j\in N(i)}\frac{e_{ji}}{c_{ji}}h^{(l)}_jW^{(l)})
\end{equation}
where $N(i)$ is the set of neighbors of node $i$, $c_{ji}$ is the product of the square root of node degrees, the superscript $l$ or $l+1$ represent the $l$-th layer or the $l+1$-th layer, $\sigma$ is an activation function, $h^{(l)}_j$ represents the input, $h^{(l+1)}_i$ represents the output, $W^{(l)}$ represents the weight matrix, and $e_{ji}$ is the scalar weight on the edge from node $j$ to node $i$.

All nodes in the neighborhood share the same convolution kernel parameters, and the performance of the model may be limited. To differentiate between nodes of different importance in the neighborhood, we can apply the attention mechanism to the graph convolution model using GAT layers, so that neighbor nodes can be given different weights and importance. DGL also gives the implementation of the GAT layer \cite{velickovic2018graph}, and the mathematical definition can be described as:
\begin{equation}
h^{(l+1)}_i=\sum_{j\in N(i)}\alpha_{i,j}W^{(l)}h^{(l)}_j.
\end{equation}
The attention score between node $i$ and node $j$ is represented as $\alpha_{ij}$:
\begin{equation}
\begin{aligned}
&\alpha^l_{ij}=softmax_i(e^l_{ij}),\\  &e^l_{ij}=LeakyReLU(\vec{a}^{T}[Wh_i||Wh_j]),
\end{aligned}
\end{equation}
where $\cdot^{T}$ represents transposition and $||$ is the concatenation operation. The attention mechanism can be defined as a single-layer feedforward neural network and can be parameterized by the weight vector $\vec{a}$. 

In addition, because there are several different types of nodes and relations, we should also aggregate information from different parts of the graph together. Therefore, we use the following method to perform the heterogeneous graph convolution:
\begin{equation}
h^{(l+1)}_{dst}=\underset{r\in R,r_{dst}=dst}{\mathrm{AVG}}(f(g_r,h^l_{r_{src}},h^l_{r_{dst}}))
\end{equation}
where $f$ is GCN or GAT module corresponding to each relation $r$, the subscripts $src$ and $dst$ represent the source and the destination of a relation $r$, and $g_r$ represents the graph containing $r$. The aggregation method used to aggregate the results from different relations in this paper is defined as average.

However, it is difficult for this WiAGCN model to use few labeled data to train a robust model. Therefore, we propose the WiDAGCN model, which can be trained by the semi-supervised method, making full use of unlabeled data. The structure of the WiDAGCN model is shown in Fig. \ref{fig:WiAGCN/WiDAGCN}. The shallow layers of WiDAGCN are the same as the WiAGCN model so we can use the pretrained model as a reliable initial condition. For the sake of brevity, we call this same part of network $G(x)$, representing the graph feature extractor. The difference is the use of the adversarial method to minimize the difference between the source and target domains, allowing the model to extract cross-domain features and improve the accuracy of the regressor while making it difficult for the domain discriminator to identify the domains that the data come from. We build a 2-layer regressor $R(x)$ and a 3-layer domain discriminator $D(x)$ with gradient reversal layer (GRL) \cite{JMLR:v17:15-239} for the adversarial model. With the help of GRL, the forward and backward propagation of the model is different:
\begin{equation}
GRL(x)=x,\ \ \frac{dGRL}{dx}=-\alpha I,
\end{equation}
where $GRL(x)$ represents the GRL function, $I$ is the identity matrix and $\alpha$ is a constant.

\subsection{Supervised Learning}
\label{sec:Supervised Learning}
For the Deep Sets model and the WiAGCN model, simple fine-tuning transfer is used. First, we use labeled source domain data to train a pretrained model:
\begin{equation}
\underset{\theta_G,\theta_R}{\mathrm{argmin}}\sum^n_{i=1}\mathrm{E}_{(x,y)\in D_{S_i}}[\mathcal{L}_y(R(G(x)),y)].
\end{equation}

Mean squared error (MSE) loss $\mathcal{L}_y$ is used in the coordinate regression problem. The $(x,y)\in D_{S_i}$ means data $x$ and label $y$ from the source domain $D_{S_i},\ i=0,1,2,...$, and $\theta$ represents the model parameters. Then, three different fine-tuning methods are compared:
\begin{equation}
\label{equ:G fine-tuning}
\underset{\theta_G}{\mathrm{argmin}}\sum^n_{i=1}\mathrm{E}_{(x,y)\in D_{T}}[\mathcal{L}_y(R(G(x)),y)],
\end{equation}
\begin{equation}
\label{equ:R fine-tuning}
\underset{\theta_R}{\mathrm{argmin}}\sum^n_{i=1}\mathrm{E}_{(x,y)\in D_{T}}[\mathcal{L}_y(R(G(x)),y)],
\end{equation}
\begin{equation}
\label{equ:All fine-tuning}
\underset{\theta_G,\theta_R}{\mathrm{argmin}}\sum^n_{i=1}\mathrm{E}_{(x,y)\in D_{T}}[\mathcal{L}_y(R(G(x)),y)].
\end{equation}

The $(x,y)\in D_{T}$ represents a small amount of labeled data $(x,y)$ in the target domain $D_{T}$. Equations \ref{equ:G fine-tuning} and \ref{equ:R fine-tuning} show the $G(x)$ fine-tuning and $R(x)$ fine-tuning, respectively. For example, in the equation \ref{equ:G fine-tuning} and \ref{equ:R fine-tuning}, we tune the $G(x)$ ($R(x)$) network and fix the parameters of the other part. For the Deep Sets model, due to the existence of the sum operation, the network is divided into two parts, the $\phi$ network, and the $\rho$ network. Similarly, we also fine-tune the $\phi$ network and the $\rho$ network separately. In addition, as shown in the equation \ref{equ:All fine-tuning}, we also perform an all-fine-tuning training, using this small amount of data to tune all the parameters of the model.

\subsection{Semi-Supervised Adversarial Domain Adaptation}
\label{sec:Semi-Supervised Adversarial Domain Adaptation}
As introduced in Section \ref{sec:Supervised Learning}, if we only have a very small amount of labeled target domain data, it may be difficult or even unable to perform the fine-tuning process. Thus, the following training framework is used based on our problem. The goal is to learn a feature extractor network that produces domain-invariant features. Such features can be learned with the discriminator $D(x)$ acting as an adversary. In this way, the generalized features can be aligned to the same distribution, no matter the data that come from which domain. The graph feature extractor $G(x)$ and regressor $R(x)$ are trained to make the $R(x)$ get accurate positioning coordinates. The point is, that the feature extractor $G(x)$ is also updated to make the discriminator $D(x)$ can not give the correct domain classification results. This competing training goal can be described like:
\begin{equation}
\label{equ:update G and R}
\begin{aligned}
&\underset{\theta_G,\theta_R}{\mathrm{argmin}}\sum^n_{i=1}\mathrm{E}_{(x,y)\in D_{S_i}}[\mathcal{L}_y(R(G(x)),y)\\&-\mathcal{L}_d(D(G(x)),d_{S_i})]
-\mathrm{E}_{x\in D^x_T}[\mathcal{L}_d(D(G(x)),d_T)],
\end{aligned}
\end{equation}
\begin{equation}
\label{equ:update D}
\begin{aligned}
\underset{\theta_D}{\mathrm{argmin}}&\sum^n_{i=1}\mathrm{E}_{(x,y)\in D_{S_i}}[\mathcal{L}_d(D(G(x)),d_{S_i})]\\
&+\mathrm{E}_{x\in D^x_T}[\mathcal{L}_d(D(G(x)),d_T)],
\end{aligned}
\end{equation}
where $\mathcal{L}_d$ is the cross-entropy loss used to calculate the loss of domain classification, $x\in D^x_T$ represents the unlabeled data in the target domain, and $d$ represents the domain label. With equation \ref{equ:update G and R}, the parameters of the feature extractor and regressor are updated so that the regressor can give accurate coordinates, while the discriminator cannot predict the correct domains. And equation \ref{equ:update D} updates the discriminator to correctly predict the domain labels. Using the GRL, we can combine these two competing steps together:
\begin{equation}
\begin{aligned}
\underset{\theta_G,\theta_R,\theta_D}{\mathrm{argmin}}\sum^n_{i=1}&\mathrm{E}_{(x,y)\in D_{S_i}}[\mathcal{L}_y(R(G(x)),y)\\
&+\mathcal{L}_d(D(GRL(G(x))),d_{S_i})]\\
&+\mathrm{E}_{x\in D^x_T}[\mathcal{L}_d(D(GRL(G(x))),d_T)].
\end{aligned}
\end{equation}

\section{Experiment}
\label{sec:Experiment}
The experiment environment in this article: The operating system is Ubuntu 18.04.5 LTS, the CPU is Intel Xeon Silver 4210 @ 2.20GHz, and the GPU is GeForce RTX 2080 Ti.
\subsection{Experiment Setup}
\label{sec:Experiment Setup}
\begin{figure*}[ht]
\centering
\subfigure[]{
    \includegraphics[height=5cm]{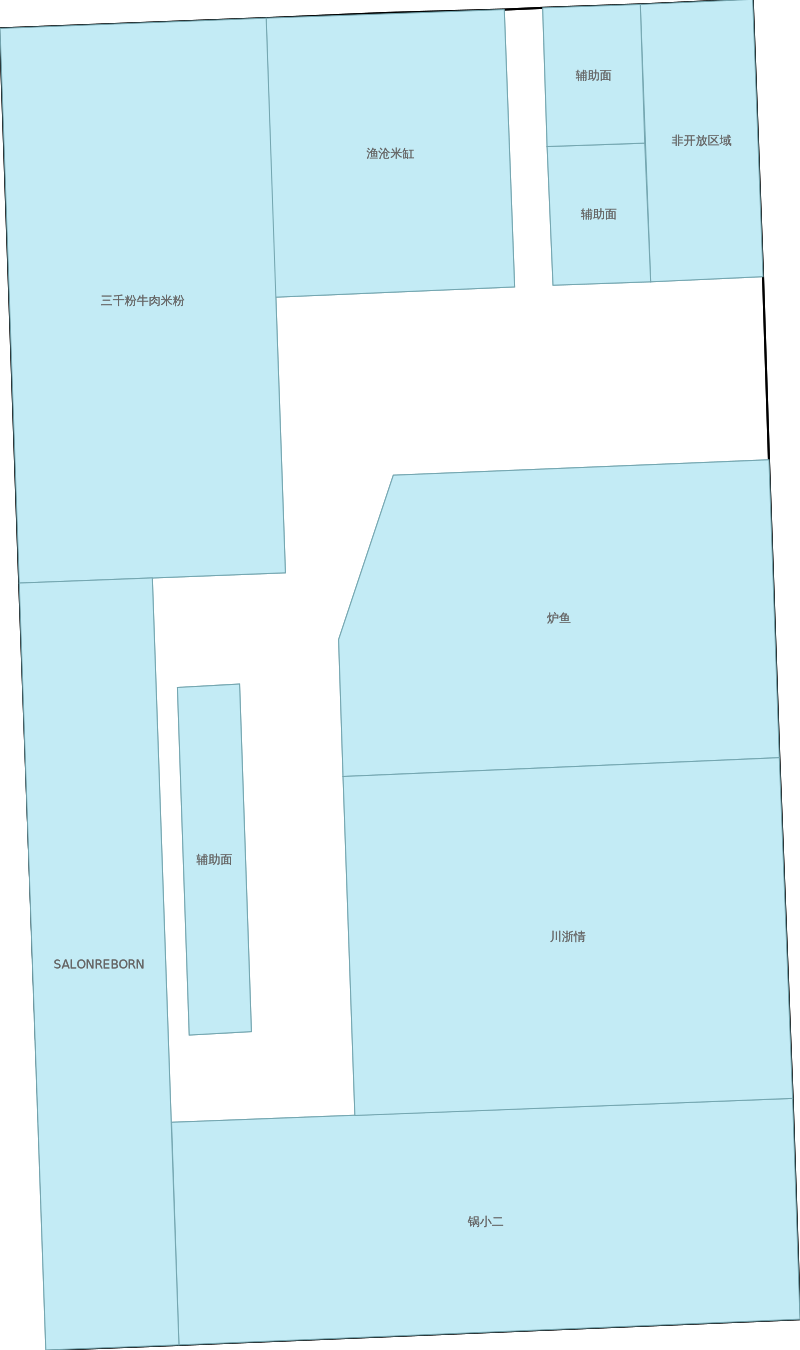}
}
\subfigure[]{
    \includegraphics[height=5cm]{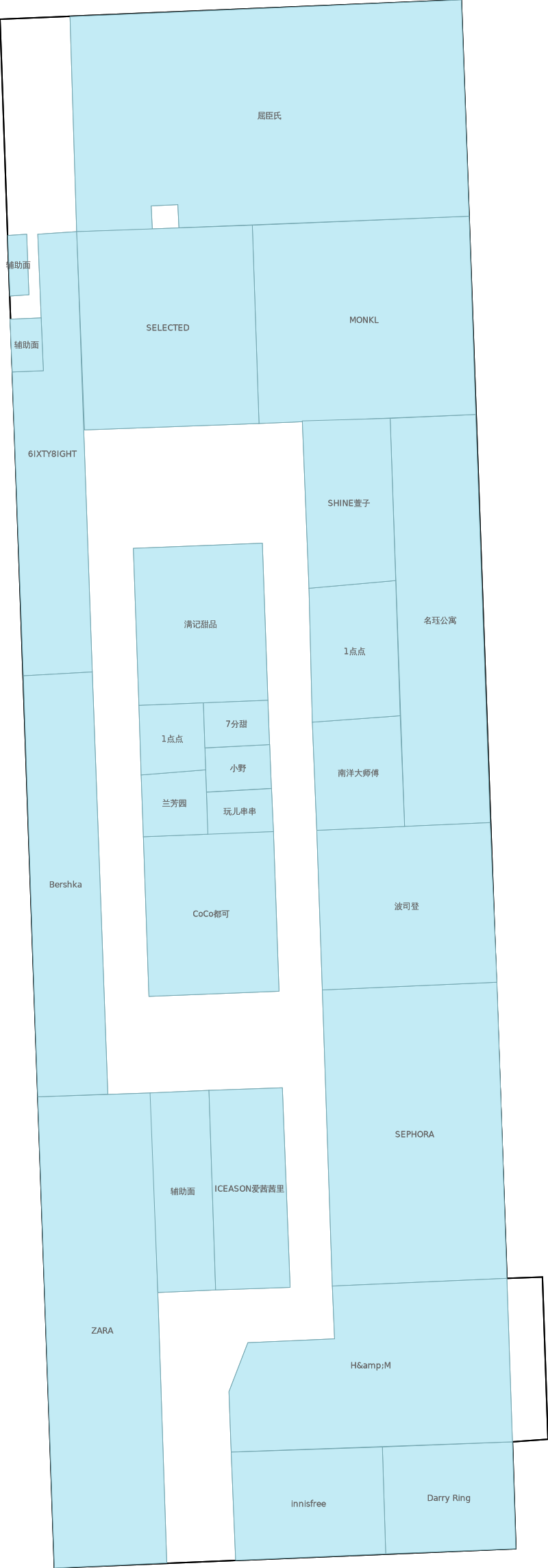}
}
\subfigure[]{
	\includegraphics[height=5cm]{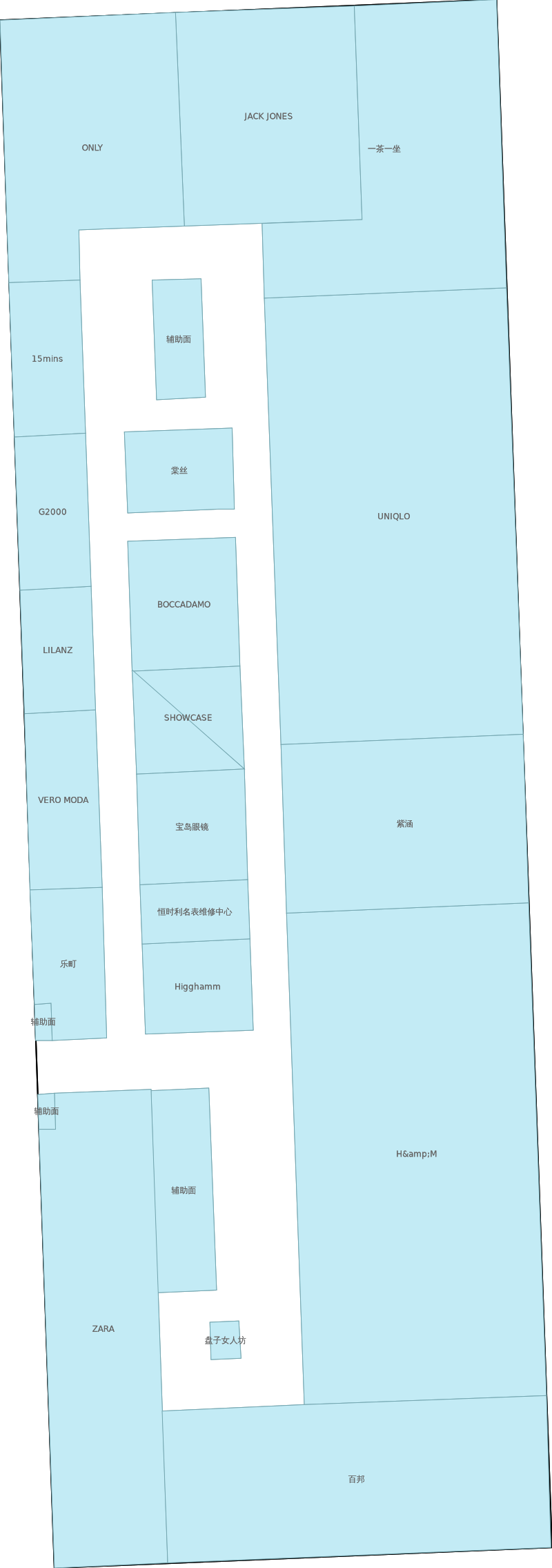}
}
\subfigure[]{
    \includegraphics[height=5cm]{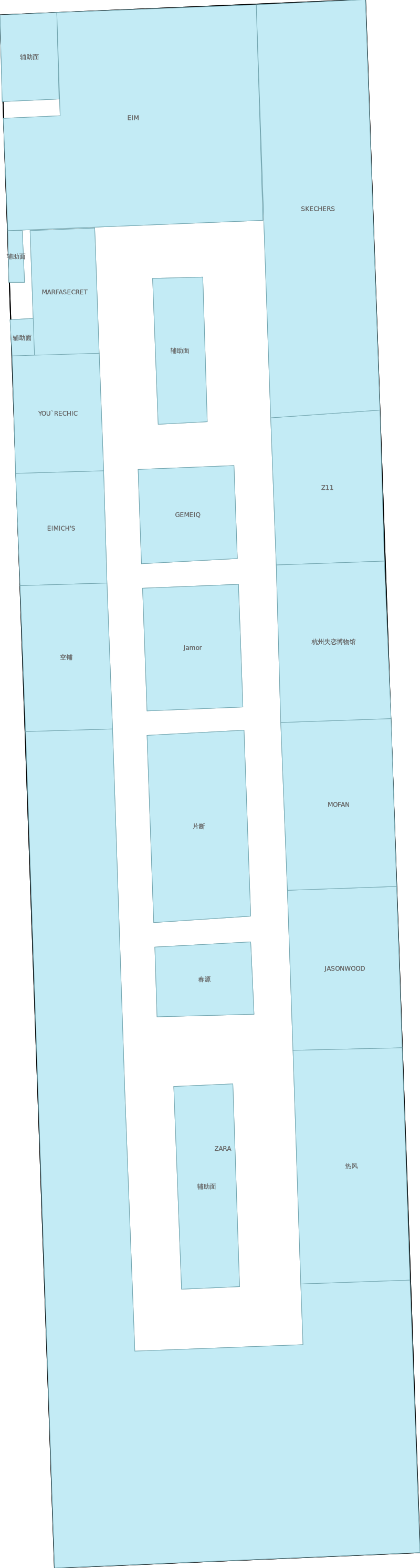}
}
\subfigure[]{
	\includegraphics[height=5cm]{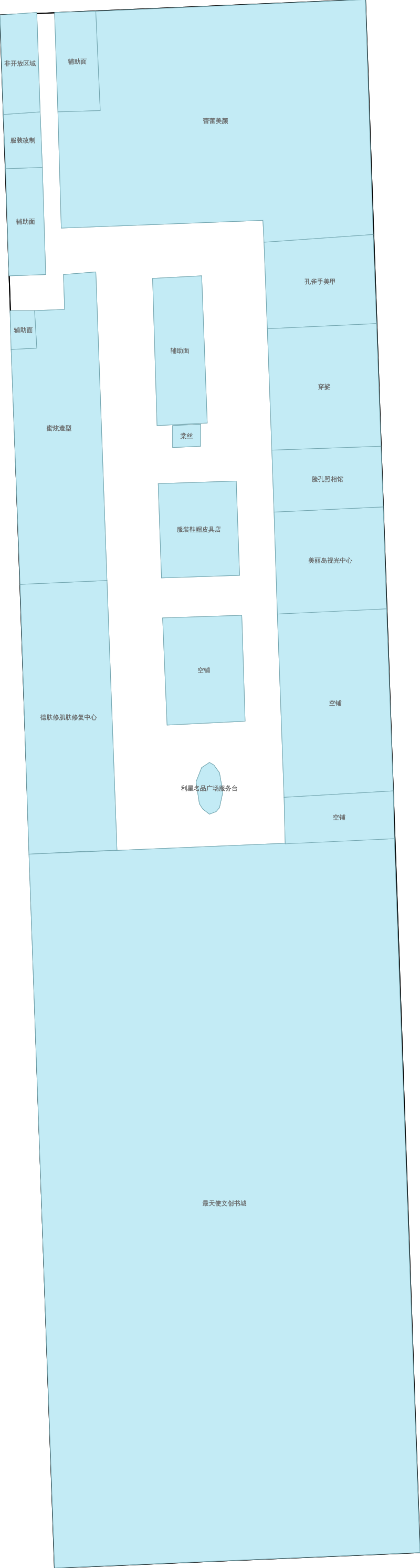}
}
\caption{The floor plans of the target site. Subfigures (a), (b), (c), (d), and (e) respectively represent floors B1, F1, F2, F3, and F4 in the target site. As shown in the figure, the structure of each floor is different, which means that the knowledge of a single floor cannot be directly used for other floors.}
\label{fig:floor_plan}
\end{figure*}

In this paper, we mainly focus on adaptation on different floors. Although floors in the same site may share many APs, different floor plans and signal attenuation can confuse the model. As mentioned in Section \ref{sec:Dataset Preparation}, site 5 is chosen as the target site, where our adaptation framework mainly works on. Details of the floor information are shown in Table \ref{tab:floor_details}. And the floor plans are also shown in the Fig. \ref{fig:floor_plan}.
\begin{table}\centering
  \caption{Detailed Floor information of the site 5.}
  \label{tab:floor_details}
  \begin{tabular}{cccc}
    \toprule
    Floor&Width (m)&Length (m)& Waypoint Number\\
    \midrule
    B1 &52.02  &87.83 &716 \\
    F1 &135.59 &47.41 &582 \\
    F2 &135.70 &47.79 &1756 \\
    F3 &135.25 &36.25 &552 \\
    F4 &135.25 &36.25 &645 \\
  \bottomrule
\end{tabular}
\end{table}

\begin{table}\centering
  \caption{Average errors (unit: m) of different fine-tuning methods when using 10\% of the labeled training data.}
  \label{tab:Fine-tuning methods}
  \begin{tabular}{lcccc}
    \toprule
    \multirow{2}{*}{Model}&\multirow{2}{*}{\makecell[c]{If use pre-\\trained model}}&\multicolumn{3}{c}{Fine-tuning Method}\\
    \cline{3-5}
    &&all-fine-tuning&$\phi$ or $G$&$\rho$ or $R$\\
    \midrule
    {Deep NN}                   &$\checkmark$   &15.72 &15.76  &21.74\\
    \midrule
    \multirow{2}{*}{Deep Sets}  &$\checkmark$   &8.22  &8.30   &7.70\\
                                &$\times$ &12.75 &12.74  &12.96\\
    \midrule
    \multirow{2}{*}{GCN}        &$\checkmark$   &6.95  &7.02   &7.41\\
                                &$\times$ &7.38 &7.44  &7.84\\
    \midrule
    \multirow{2}{*}{\makecell[l]{WiAGCN\\(Proposed)}}  &$\checkmark$   &\textbf{6.58}  &\textbf{6.61}   &\textbf{7.00}\\
                                &$\times$ &6.70 &6.69  &7.42\\
  \bottomrule
\end{tabular}
\end{table}

\begin{table*}\centering
  \caption{Impact of labeled training set size on average errors (unit: m)}
  \label{tab:Positioning errors}
  \begin{tabular}{lccccccccc}
    \toprule
    \multirow{2}{*}{Model}&\multirow{2}{*}{Pretraining}&\multirow{2}{*}{\makecell[c]{Permutation\\Invariant Operation}}&\multirow{2}{*}{\makecell[c]{Adversarial\\Component}}&\multicolumn{6}{c}{Size of Labeled Training Dataset} \\
    \cline{5-10}
    &&&&1\%&5\%&10\%&20\%&30\%&40\%\\
    \midrule
    KNN &$\times$ &$\times$ &$\times$ &25.46  &22.87   &22.49   &22.17   &21.91   &21.45   \\
    \midrule
    Decision Tree&$\times$ &$\times$ &$\times$&24.84 &16.41   &14.63   &12.51   &11.42   &11.06   \\
    \midrule
    Deep NN&$\checkmark$ &$\times$ &$\times$&21.45 &17.64   &15.72   &13.74   &12.71   &11.93   \\
    \midrule
    SAE&$\checkmark$ &$\times$ &$\times$&51.00 &21.67 &16.26 &13.14 &12.15 &11.10   \\
    \midrule
    1D-CNN&$\checkmark$ &$\times$ &$\times$&21.48 &16.30 &11.99 &10.93 &9.90 &9.53   \\
    \midrule
    Deep Sets&$\checkmark$ &$\checkmark$ &$\times$&13.45  &9.39   &8.22   &6.03   &5.61   &5.62   \\
    \midrule
    DANN&$\checkmark$ &$\times$ &$\checkmark$&19.47  &11.28   &8.91   &6.53   &5.36   &5.09   \\
    \midrule
    GCN&$\checkmark$ &$\checkmark$ &$\times$&10.05  &7.80   &6.95   &6.22   &5.83   &5.56   \\
    \midrule
    \makecell[l]{WiAGCN\\(Proposed)}&$\checkmark$ &$\checkmark$ &$\times$   &9.67   &\textbf{7.35}   &6.58   &5.97   &5.58   &5.32   \\
    \midrule
    \makecell[l]{WiDAGCN\\(Proposed)}&$\checkmark$ &$\checkmark$ &$\checkmark$&\textbf{9.09}   &7.59   &\textbf{6.04}		&\textbf{5.67}   &\textbf{5.20}   &\textbf{4.84}   \\
  \bottomrule
\end{tabular}
\end{table*}

\begin{table}\centering
  \caption{Performance of models with different scales (unit: m)}
  \label{tab:HiddenLayerSize}
  \begin{tabular}{p{2cm}<{\centering}p{1cm}<{\centering}p{1cm}<{\centering}p{1cm}<{\centering}}
    \toprule
    \multirow{2}{*}{\makecell[c]{{Hidden Layer}\\Size}}&\multicolumn{3}{c}{Size of Labeled Training Dataset} \\
    \cline{2-4}
    &20\%&30\%&40\%\\
    \midrule
    16   &5.77	&5.18	&4.92   \\
    \midrule
    32   &5.67	&5.20	&4.84   \\
    \midrule
    64   &6.00	    &5.30	&5.06   \\
  \bottomrule
\end{tabular}
\end{table}

These 5 floors are set as the target domain in turn, while the other 4 floors are set as 4 source domains. The labels of source domain data are all known, and some of the target domain labels are known. We test the model on conditions of different sizes (1\%, 5\%, 10\%, 20\%, 30\%, 40\%) of training data with known labels. In the supervised learning setting, the labeled data in the target domain are used for training, and the remaining unlabeled data are divided into a validation set and a test set according to a 2:8 ratio. In the adversarial training setting, all data in the target domain participate in the training process, regardless of whether they are labeled. And the validation set and test set are also divided according to a ratio of 2: 8 with the unlabeled data. Thus, the testing and validation process can also be viewed as using the model to label the unlabeled data. For the adversarial model, let these labeled waypoints as $(x,y)\in D_{T}$ and the rest unlabeled fingerprints as $x\in D^x_T$. The initial learning rate of $G$, $C$ and $R$ are all set to 0.001, and if the validation loss does not decrease for 10 consecutive epochs, the learning rate of $G$ and $R$ will be halved. All $G$, $C$ and $R$ networks use Adam optimizer. For the GRL, the initial value of $\alpha$ is set to 0, and the value of $\alpha$ increases by 0.0001 with each epoch. The total number of the training epoch is set as 250 for each experiment. Ten times repeating experiments are performed for each floor, so the result of each model for each training data proportion is the average of total 50 times experiment results. As illustrated in Fig. \ref{fig:floor_plan}, the structures of the floors in the target site vary from one another. Additionally, it is feasible for APs located on a single floor to be detected by wireless devices situated on multiple floors. Consequently, even though the RSSIs of a specific AP can be utilized for localization purposes across different floors, the propagation of WiFi signals is susceptible to alterations between floors due to the varying floor heights and the influences of walls, ceilings, and other environmental factors. By conducting the experimental design described above, we aim to allow the model to learn general knowledge across the different floors. This knowledge could be in the form of a mapping of data distribution in a high-dimensional space, where data from different domains can be aligned. Our goal is to achieve a more accurate indoor positioning system by leveraging this knowledge. This would involve learning how the WiFi signals are affected by the structure and other factors of each floor, and how they are transmitted between floors, so that the model can make more accurate predictions about a user's location.

\subsection{Results}
The total time for constructing the subgraph in the target site was 1926 seconds, or approximately 32 minutes. As there were 5 floors in the target site and 5 repetitions of sampling were conducted on each floor, the average time cost of constructing the subgraph on a single floor was approximately 77.04 seconds. In total, the 50 repeated experiments took approximately 100 hours to complete, with each experiment taking approximately 2 hours. During the experiment process, the GPU memory usage ranged from 1.6GB to 5.0GB. While the resources consumed during testing are of particular concern for the practical application of the model, our model was able to generate approximately 430 localization results in 11 seconds, with each localization process taking approximately 0.026 seconds. The GPU memory usage during testing was approximately 0.9GB, including the test dataset, models, and other consumption.

We first compare the performance of different fine-tuning methods of supervised models. As we mentioned previously, we fine-tune the $\rho$ or $G$ network, $\phi$ or $R$ network and the whole network, respectively. And we also compare the accuracy of models that use or do not use other sites to pretrain. The average positioning errors of 50 repeated experiments are shown in Table \ref{tab:Fine-tuning methods}. Because semi-supervised WiDAGCN does not apply to the fine-tuning method, it is not concluded in Table \ref{tab:Fine-tuning methods}. The performance of WiDAGCN will be described in more detail later.

As mentioned in Section \ref{sec:Deep Sets}, a Deep NN model consisting of only fully-connected layers without permutation invariant operation is built. Similarly, a GCN model without an attention mechanism and multilevel feature aggregation is also implemented. Despite having similar structures, the Deep NN model performed poorly compared to the Deep Sets network. By using the permutation invariant operation, the positioning error was reduced by 47.70\%, 47.33\%, and 64.58\% in the cases of all-fine-tuning, $\phi$ fine-tuning, and $\rho$ fine-tuning, respectively. The results showed that fine-tuning the deep layers of the Deep Sets model outperformed fine-tuning the shallow layers, and there was no significant difference between all-fine-tuning and $\phi$ network fine-tuning. This finding is consistent with the results presented by J. Yosinski et al. \cite{yosinski2014transferable}, which showed that shallow-layer features are applicable to different tasks and are commonly kept fixed in transfer learning while the deep layers are updated according to the new dataset. Both of the tested graph-based models showed better performance with only a few dozen labeled data available. On the one hand, GCN has the property of maintaining a good performance even if data are sparse \cite{DBLP:conf/iclr/KipfW17}. On the other hand, higher-order neighbors are also considered so that more complex features can be utilized. However, the graph-based models showed some inconsistent results. Unlike Deep Sets, shallow layers fine-tuning performed better in terms of accuracy for GCN. The difference between all-fine-tuning and $G$ network (corresponding to $\phi$ network in Deep Sets) fine-tuning was also not significant. There has been limited research on fine-tuning GCN, but we can find some similarities in a study of CNN. M. Amiri et al. found that shallow layers fine-tuning outperformed deep layers fine-tuning when using a small dataset with U-Net \cite{amiri2020fine}. In our problem, this may be due to low-level graph patterns associated with shallow layers.

The results of using different sizes of the labeled training dataset are shown in Table \ref{tab:Positioning errors}. We have marked several operations conducted on compared models to conveniently compare the effect of different components of the model. To further validate the improvement of the adversarial component, we also built a domain-adversarial neural network (DANN) \cite{JMLR:v17:15-239} with a similar structure to the Deep NN and Deep Sets models. In addition, we implemented two previous indoor localization systems, SAE \cite{belmannoubi2019deep} and 1D-CNN \cite{hsieh2019deep}, in our environment. In the original study of SAE, the researchers trained the SAE in an unsupervised way like the common autoencoder \cite{belmannoubi2019deep}. The input and output were both the fingerprint vector, and each layer was trained independently to extract high-level features. Then, the encoder part was connected to a classifier and the decoder part was discarded. Finally, a supervised training process was conducted to obtain the localization model. We followed the steps and parameter setup of the original study and applied the model to our environment. We also implemented the 1D-CNN following the original study. The result shows that both the permutation invariant operation and the adversarial component can help the model obtain a higher accuracy from the comparison between the Deep NN, Deep Sets and DANN, and both the Deep Sets and the DANN models were able to predict positions more accurately than the SAE and 1D-CNN models in an environment with limited labels. This demonstrates that both the permutation invariant operation and the adversarial component can help improve the performance of the model. The DANN performed poorly when the number of labeled data was extremely limited, but its performance improved significantly as the number of labels increased. The results of the WiAGCN and the WiDAGCN models also showed that the adversarial component had a positive effect on the graph-based model. When 40\% labeled data (220\textasciitilde702 pieces of data) are used, the model achieves the best improvement of 9.02\% (from 5.32m to 4.84m). The graph-based models obtain better accuracy when there are only 1\% or 5\% labels. Our proposed WiDAGCN model combines the advantages of above components and achieves the best results overall. WiDAGCN performed slightly worse than WiAGCN in the case of 5\%, which could be related to the distribution of the data at the time of sampling.

To delve deeper into the performance of the proposed WiDAGCN model, additional experiments were conducted using models of varying scales. In the above experiments, the size of the GCN hidden layer in the WiDAGCN model was set to 32 for each layer. Subsequently, the performance of models with hidden layer sizes of 16 and 64 was evaluated, using the same set of hyperparameters, to analyze the impact of the scale of the model on performance. The results of these experiments are presented in Table \ref{tab:HiddenLayerSize}. In terms of training time, there is no significant difference between the three models. Moreover, the difference in localization accuracy is also not noticeable, but the model with hidden layer size of 32 achieved better results. This is likely due to the fact that larger models still have not converged completely with the same hyperparameter settings, while smaller models are somewhat limited in their expressive capabilities. In this way, we can strike a balance between model complexity and expressiveness, using the simplest and most efficient models to address practical problems.

\section{Discussion}
\label{sec:Discussion}
In this study, our primary goal was to improve localization accuracy, taking into account the complexity and changes of the environment. In comparison to some existing studies, we focused on the approach of using fewer labeled data to address the problem in more complex application environments, particularly large-scale multi-floor buildings. For example, the ViVi model \cite{wu2017gain} performed well in three test areas: an office building, an academic building, and a classroom building. However, tens of thousands of pieces of data were used for model training. The weighted algorithm \cite{8362651} used only 2680 pieces of data, but the experiment environment was much simpler than the application environment. Our scenarios were set in a shopping mall with 5 floors, which is several times larger than other common experiment environments. In summary, we were able to achieve localization accuracy similar to existing studies using a small amount of labeled data in more complex, multi-environment settings.

There are still some issues that need to be addressed or that we have overlooked in this study. We believe that constructing a more rational graph structure is key to solving the problem. The construction of graphs for indoor positioning problems is a topic that merits further discussion, and how to construct a graph with a simple structure that better reflects the original information is also a topic we will investigate in the future. From a model design perspective, the subgraph regression approach used in this study focuses on the nodes and all parts within their second-order neighbors. In terms of maintaining the permutation invariance of features, the subgraph regression approach is more suitable for the Deep Sets model. However, in terms of the localization problem itself, it is more akin to a node regression problem, where the model needs to give the coordinate values corresponding to the nodes. The advantages and disadvantages of these two approaches need to be analyzed and compared further.

\section{Conclusion}
\label{sec:Conclusion}
In this paper, we propose an RSSI-based GCN indoor localization model called WiDAGCN. We construct heterogeneous graphs based on the signal connections between users' devices and APs in order to consider information from neighbors or higher-order neighbors, or even unlabeled nodes. We design a domain adversarial GCN model to align the distribution of data from different environments. We focus on permutation invariant features and obtain the aggregation representation of the graph. We test our model using a novel, real-world indoor localization public dataset and find that it achieves competitive performance compared to state-of-the-art methods. This means that we can conduct simple site surveys and use unlabeled signal records received by user devices as auxiliary data to train a reliable positioning model.

\bibliographystyle{IEEEtran}
\bibliography{ref}


\end{document}